\documentclass[acmsmall]{acmart}
\AtBeginDocument{%
  }

\setcopyright{cc}
\setcctype{by}
\acmDOI{10.1145/3832243}
\acmYear{2026}
\acmJournal{PACMSE}
\acmVolume{3}
\acmNumber{ISSTA}
\acmArticle{ISSTA152}
\acmMonth{10}
\acmSubmissionID{issta26main-p1636-p}
\received{2026-01-30}
\received[accepted]{2026-06-25}

\usepackage[utf8]{inputenc}
\usepackage{multirow}
\usepackage{tabularx}
\usepackage{algorithm}
\usepackage{algpseudocode}
\usepackage[many]{tcolorbox}
\tcbset{
  boxsep=1pt,
  left=6pt,
  right=6pt,
  top=6pt,
  bottom=6pt,
  before skip=7pt,
  after skip=7pt
}
\usepackage{amsmath}
\usepackage{xspace}
\usepackage{siunitx}

\newcommand{\tool}{\textsc{FCC}\xspace}
\newcommand{\kernjc}{\textsc{KernJC}\xspace}
\newcommand{\kconfig}{\textsc{Kconfig}\xspace}
\newcommand{\kbuild}{\textsc{Kbuild}\xspace}
\newcommand{\nvd}{\textsc{NVD}\xspace}
\newcommand{\syzbot}{\textsc{syzbot}\xspace}
\newcommand{\kasan}{\textsc{KASAN}\xspace}
\newcommand{\kmsan}{\textsc{KMSAN}\xspace}
\newcommand{\makeolddef}{\texttt{make olddefconfig}\xspace}
\newcommand{\olddef}{\texttt{olddefconfig}\xspace}
\newcommand{\makedef}{\texttt{make defconfig}\xspace}
\newcommand{\configfile}{\texttt{.config}\xspace}
\begin{document}


\title{Inferring 1-Minimal Trigger Configurations for Assessing Linux Kernel CVE Triggerability}

\author{Tongjie Wei}
\orcid{0009-0004-5766-9677}
\affiliation{%
  \institution{Nanjing University of Science and Technology}
  \department{School of Cyber Science and Engineering}
  \city{Nanjing}
  \country{China}
}
\email{weitongjie@njust.edu.cn}

\author{Peng Zhang}
\authornote{Peng Zhang and Gangyan Zeng are the corresponding authors.}
\orcid{0000-0001-9518-5914}
\affiliation{%
  \institution{Nanjing University of Science and Technology}
  \department{School of Cyber Science and Engineering}
  \city{Nanjing}
  \country{China}
}
\email{zhang\_peng@njust.edu.cn}

\author{Zhiwen Hu}
\orcid{0009-0004-8766-4762}
\affiliation{%
  \institution{Nanjing University of Science and Technology}
  \department{School of Cyber Science and Engineering}
  \city{Nanjing}
  \country{China}
}
\email{huzhiwen@njust.edu.cn}

\author{Xupu Hu}
\orcid{0009-0002-8896-1203}
\affiliation{%
  \institution{Nanjing University of Science and Technology}
  \department{School of Cyber Science and Engineering}
  \city{Nanjing}
  \country{China}
}
\email{huxupu@njust.edu.cn}

\author{Chen Lyu}
\orcid{0000-0002-5044-1459}
\affiliation{%
  \institution{Shandong Normal University}
  \city{Jinan}
  \country{China}
}
\email{lvchen@sdnu.edu.cn}

\author{Gangyan Zeng}
\authornotemark[1]
\orcid{0000-0003-2696-8549}
\affiliation{%
  \institution{Nanjing University of Science and Technology}
  \department{School of Cyber Science and Engineering}
  \city{Nanjing}
  \country{China}
}
\email{gyzeng@njust.edu.cn}

\begin{abstract}
Vendors assessing Linux kernel CVEs need to know whether a bug is triggerable under production-tailored configurations, not merely whether a version is affected, yet upstream reproducers and vulnerability databases rarely provide configuration-level context.
We study minimal trigger-configuration inference: given a CVE entry and a target kernel version (optionally a baseline \configfile), we synthesize a \kconfig-satisfiable option set that remains effective after \makeolddef and, when a reproducer is available, still triggers under a specified evaluation protocol; we then prune it to a 1-minimal (subset-minimal) boundary for evaluation.
Our framework \tool links vulnerability cues to build-system symbols, completes implicit prerequisites under \olddef feedback to avoid silent rollback, and performs runtime-validated minimization guided by dependency topology.
We evaluate on \kernjc and KernelCTF, totaling 88 CVEs across multiple kernel versions.
On the 88-CVE set, \tool improves the post-\makeolddef configuration success rate from 62.5\% (55/88) to 96.6\% (85/88) over an olddef-only injection baseline; on the \kernjc set, \tool reduces the average candidate set size by 78.7\% compared to \kernjc (Avg.\ 14.72 vs.\ 69.00 options per CVE). 
A stage-wise analysis of time and token costs shows that Stage~I dominates overhead, while CVE-focused evidence selection substantially reduces this cost.
By returning an effective and auditable 1-minimal configuration boundary, \tool helps vendors scope triggerability against their deployment configurations with a clear, tool-supported decision line.
\end{abstract}




\maketitle

\section{Introduction}
For Linux kernel vulnerability risk assessment, vendors often care less about whether a kernel version is \emph{affected} and more about whether a vulnerability is \emph{triggerable} under production-tailored kernel configurations~\cite{Zou2024NDSS}.
Many CVEs manifest only under specific configuration combinations~\cite{AiushiUSENIX22,Chen2021Trustcom,Ruan2024RAID}, and there is a substantial configuration gap between upstream reproduction environments and product kernels.
To address this gap, we study \emph{minimal trigger-configuration inference} for vulnerability reproduction.
Given a CVE entry, a target kernel version, and optionally a baseline \configfile and a runnable proof-of-concept (PoC), we synthesize a triggering option set that (i) satisfies \kconfig constraints, (ii) remains effective after \makeolddef\footnotemark%
\footnotetext{The command \texttt{make olddefconfig} updates an existing kernel configuration for a target version by assigning default values to newly introduced options and resolving Kconfig constraints; options with unsatisfied prerequisites may be silently unset.}
, and (iii) triggers the vulnerability under an oracle-defined evaluation protocol (Sec.~\ref{sec:oracles}) validated in our upstream QEMU/KVM setting.
We then prune it to a 1-minimal (subset-minimal) triggering option set that serves as an actionable evaluation boundary for production deployments; without a runnable PoC, we only output an effective configuration boundary.
In practice, fuzzing-oriented platforms such as \syzbot~\cite{google2026syzbot} often reproduce bugs under feature-rich, coverage-oriented kernel configurations, whereas production kernels are typically tailored for specific workloads and devices~\cite{Hasanov25ICSE}.

For deployments that rely on a bounded set of stock cloud or distribution images, such as Google Cloud public OS images~\cite{google_cloud_images}, triggerability can often be assessed by running the available PoC on those images. This work focuses on the more difficult setting of vendor- and downstream-tailored kernels, where maintainers must determine whether an upstream CVE remains triggerable under the deployed effective configuration, since upstream reproducibility does not necessarily imply downstream triggerability~\cite{Zou2024NDSS,Zou22USENIX,AiushiUSENIX22,Hasanov25ICSE}.
Moreover, mainstream vulnerability databases such as the \nvd~\cite{NVD2026} typically report affected kernel version ranges but rarely provide configuration-level context~\cite{Mu2018USENIX,Chen2021Trustcom,Ruan2024RAID}, making it difficult to assess triggerability in real-world production settings.

Upstream reproducers often trigger bugs under coverage-oriented configurations, yet they rarely expose (1) which configuration options are necessary to reach and trigger the vulnerable code path and (2) whether these options can be pruned into a small, auditable boundary while preserving matched trigger evidence.
Static dependency-based configuration inference helps reduce manual search, but dependency expansion alone is ill-suited for cross-subsystem runtime context and for dependency drift across kernel versions.
Our focus is complementary to reproducer adaptation across kernels/environments: we keep the reproducer fixed and target the missing configuration prerequisites that must remain effective after \makeolddef.
Therefore, turning ``triggerable under a production configuration'' into an actionable and auditable conclusion calls for an end-to-end pipeline that (a) recovers functional prerequisites from vulnerability information, (b) synthesizes a buildable configuration that remains effective under \kconfig constraints, and (c) removes redundancy under dynamic trigger evidence until converging to a 1-minimal evaluation boundary under the protocol.

Three practical issues make this pipeline non-trivial.
First, configurations determine not only whether code is compiled, but also whether it becomes reachable at runtime; trigger chains are often scattered across subsystems, so relying on local static clues alone can miss prerequisites or introduce substantial over-approximation.
Second, due to \kconfig constraints, a configuration is not effective simply because an option is written in \configfile: critical options may be silently rolled back during \makeolddef when prerequisites are unsatisfied, causing reproduction to fail even when options appear enabled.
Third, the configuration space grows exponentially with the number of features, while dependencies and implementations evolve across versions; minimizing configurations at a controlled cost without losing matched trigger evidence is therefore challenging.

Recent work has started to automate the identification of configuration options relevant to kernel vulnerabilities.
A representative approach, \kernjc~\cite{Ruan2024RAID}, infers vulnerability-related options by constructing a \kconfig dependency graph, thereby reducing manual effort.
However, static dependency expansion alone is insufficient for reproduction-oriented inference of a minimal triggering boundary under \makeolddef constraint resolution and runtime trigger evidence.
Specifically, it faces three limitations:
(1) \textbf{over-approximation}---the inferred set often contains many options irrelevant to the runtime triggering path, making it hard to derive an actionable boundary;
(2) \textbf{version sensitivity}---\kconfig dependencies evolve substantially across kernel versions, so static graphs do not generalize reliably; and
(3) \textbf{no validated effectiveness after \makeolddef}---even when correct candidates are identified, explicitly enabling them may still be overridden during \makeolddef due to constraints such as \texttt{depends on} and \texttt{select}, leading to the practical failure mode of ``correct identification but failed reproduction.''
These limitations motivate three core challenges for triggerability evaluation under production configurations.

\textit{Challenge 1: Identifying implicit runtime-context prerequisites along cross-subsystem trigger chains.}
Critical configuration prerequisites are often scattered across drivers, architectures, and core subsystems.
Relying solely on build-level static dependency graphs can miss runtime preconditions required by the trigger chain, or introduce many irrelevant options that inflate the candidate set.

\textit{Challenge 2: \kconfig-constrained configuration synthesis with validated effectiveness after \makeolddef.}
\kconfig is a logical system with implicit prerequisites; enabling an option directly in \configfile does not guarantee that it remains enabled in the final configuration.
During \makeolddef, the build system may revert or override options according to dependency relations, disabling critical switches and breaking reproduction.

\textit{Challenge 3: Denoising/minimizing configurations under kernel version evolution.}
Static inference inevitably produces an over-approximate set, while dependencies drift across versions through additions, removals, and refactoring, invalidating previously inferred information in the target version.
Naive option-by-option pruning easily leads to combinatorial explosion.
The key question is how to denoise efficiently using dynamic trigger evidence while leveraging dependency topology to constrain the search and converge to a 1-minimal triggering boundary at a controllable cost.

We present \tool (FindCveConfig), an automated framework that infers a minimal kernel configuration boundary for triggering Linux kernel vulnerabilities under an oracle-defined protocol.
\tool targets determining whether a vulnerability is triggerable for a specified kernel version and configuration context in our upstream QEMU/KVM setting, and provides a tool-supported workflow for customized production environments.
Importantly, \tool uses the LLM only for Stage~I candidate identification, Stage~II constraint parsing, and Stage~III trigger-evidence matching; retained options are still validated by \makeolddef-based effectiveness and build/run checks.

Concretely, \tool addresses the three challenges as follows.
For Challenge~1, \tool builds a semantic profile by jointly analyzing the CVE description in \nvd, the corresponding CWE information, and external resources referenced in disclosures, and uses it to guide targeted mining over \kconfig, Makefile, and other build specifications to capture dispersed prerequisites and reduce over-approximation.
For Challenge~2, \tool models \kconfig constraints, performs dependency completion and consistency checking, and validates buildability so that critical options remain effective after \makeolddef.
For Challenge~3, \tool performs dependency-topology-guided minimization under dynamic trigger evidence, constraining the search order and pruning strategy so that the candidate set empirically converges to a 1-minimal triggering boundary for the target version within a controlled time budget.

We evaluate \tool on 88 historical Linux kernel CVEs spanning multiple versions (32 from \kernjc and 56 from KernelCTF), comparing against \kernjc and an olddef-only injection baseline.
On the \kernjc set, \tool maintains full CVE coverage while reducing redundancy, decreasing the average candidate set size from 69.00 to 14.72 options per CVE (RQ1; Table~\ref{tab:rq1-coverage-size}).
Compared with the olddef-only baseline, \tool improves the post-\makeolddef configuration success rate from 62.5\% (55/88) to 96.6\% (85/88) (RQ2; Fig.~\ref{rq2_fig1}).
On the 32 \kernjc CVEs with runnable PoCs, \tool further reduces the Stage~I candidate set from 14.72 to a final boundary of 1.63 options per CVE on average while preserving matched trigger evidence under the same protocol (RQ3; Table~\ref{tab:rq3}). For RQ4, we profile per-stage time/token costs (Fig.~\ref{rq4}).

In summary, this paper makes the following major contributions.
\begin{itemize}
  \item We introduce and formalize the task of minimal trigger-configuration inference for vulnerability reproduction, and motivate its role in production risk assessment.
  \item We propose \tool, which performs vulnerability-semantics-driven candidate mining over build specifications (e.g., \kconfig and Makefile) to recover triggering prerequisites beyond static dependency expansion; 
  the LLM is used only for candidate identification, constraint parsing, and trigger-evidence matching, while all options are validated by \makeolddef-based effectiveness checks and build/run validation.
  \item We design \kconfig-constrained dependency completion with build validation to ensure that critical options remain effective after \makeolddef.
  \item We propose a dependency-topology-guided runtime minimization algorithm that yields a 1-minimal triggering boundary under the oracle-defined protocol, and validate it on the KernJC dataset with \kernjc and an olddef-only baseline.
\end{itemize}

\section{Motivation}

\subsection{A Motivating Example}
\label{sec:example}

\begin{figure}[t]
  \centering
  \includegraphics[width=\textwidth]{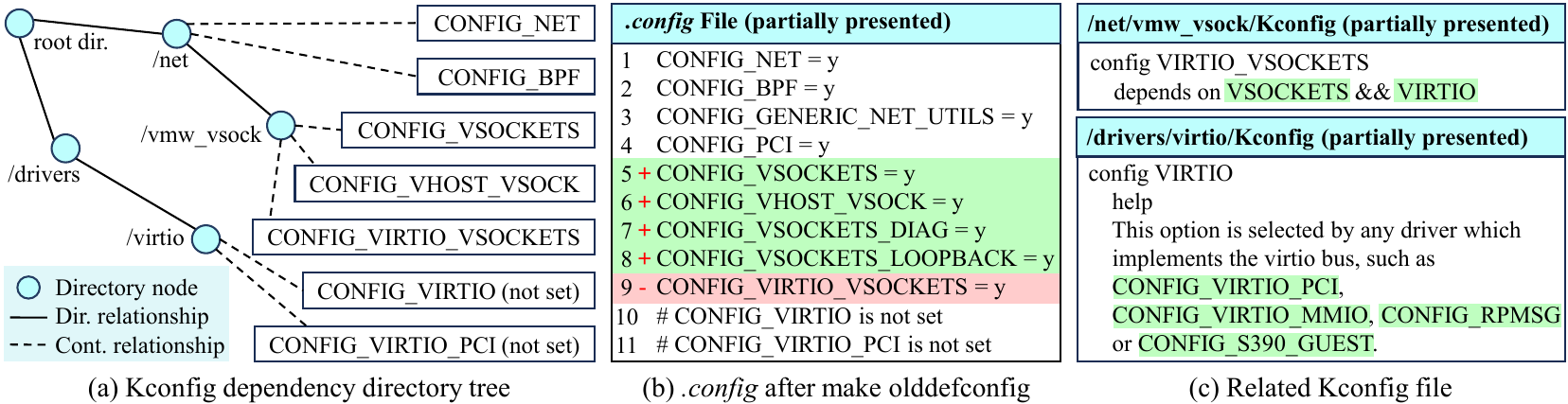}
  \caption{Motivating failure mode: ``candidate set'' $\neq$ effective, triggerable configuration under \makeolddef. (a) Relevant \kconfig constraints for CVE-2021-26708. (b) After injection, \makeolddef may silently roll back options (``+'' kept enabled; ``-'' disabled), breaking PoC triggering. (c) Root cause: some prerequisites are enabled only through feasible \texttt{select}-mediated paths (and conditional contexts), so forcing a dependent option in \configfile is insufficient.}
  \label{motivation}
\end{figure}

Figure~\ref{motivation} shows why a ``vulnerability-related'' option list can still fail to produce an effective, triggerable configuration once \makeolddef resolves \kconfig constraints.
We use CVE-2021-26708 (net/vmw\_vsock) as an example; Fig.~\ref{motivation}(a) highlights the relevant constraints.

Many dependency-based inference pipelines (with \kernjc as a representative) generate a candidate set from code locations and \kconfig relations, inject it into \configfile, and run \makeolddef to obtain a buildable configuration.
However, candidates written as enabled may not survive \makeolddef: Fig.~\ref{motivation}(b) shows silent rollback (``+'' kept; ``-'' disabled) despite explicit assignments.
In this case, \texttt{CONFIG\_VIRTIO\_VSOCKETS} is rolled back by \makeolddef, and the PoC stops producing the same trigger signature under our protocol.

The root cause is reflected in Fig.~\ref{motivation}(c).
Besides depending on \texttt{VSOCKETS}, \texttt{VIRTIO\_VSOCKETS} also depends on \texttt{VIRTIO}.
In the target version/configuration context, \texttt{VIRTIO} may not be directly user-selectable (e.g., no prompt) and is typically enabled via \texttt{select} from drivers such as \texttt{VIRTIO\_PCI}, \texttt{VIRTIO\_MMIO}, \texttt{RPMSG}, and \texttt{S390\_GUEST}.
Thus, forcing \texttt{VIRTIO\_VSOCKETS=y} is not enough; keeping it enabled requires enabling at least one feasible selector path that turns on \texttt{VIRTIO}.
This \makeolddef rollback is a recurring failure mode of one-shot injection baselines, which we quantify later under the olddef-only baseline (RQ2).

Overall, we need to (i) ensure trigger-critical options remain effective after \makeolddef and (ii) remove redundancy to obtain a small, auditable boundary.

\subsection{Key Ideas}
The example suggests three requirements: (1) map vulnerability evidence to build-system symbols; (2) ensure effectiveness under \makeolddef; and (3) minimize using runtime evidence.
Concretely, Stage~I builds a candidate set $C$, Stage~II repairs it into an effective set $S$ that survives \makeolddef, and Stage~III prunes $S$ to a 1-minimal boundary $S_{\min}$ under the trigger oracle.

Vulnerability artifacts (e.g., \nvd reports, patches, PoCs, and references) are unstructured and rarely map directly to \texttt{CONFIG\_*} symbols.
We combine \emph{build-evidence cues} (e.g., file paths, module/interface names, macros) with \emph{semantic cues} (e.g., triggering mechanism, subsystem, preconditions) to seed and constrain candidate discovery, reducing over-approximation early.

Trigger-critical switches must remain enabled after \makeolddef.
This requires handling forward prerequisites (\texttt{depends on}), \texttt{select}-mediated enabling paths, and conditional contexts (\texttt{if}/\texttt{source}).
For non-user-selectable symbols, at least one feasible enabling branch must be completed; we iteratively repair missing prerequisites under \makeolddef feedback until options become stably effective or are deemed unsatisfiable under bounded search.

Since static dependencies cannot certify necessity, we minimize via a \emph{disable--build--run} loop and accept a removal only if the configuration stays effective and the run matches the same bug signature under the protocol.

\section{Preliminaries}

\subsection{Task and Goal}
We study \emph{minimal trigger-configuration inference} for Linux kernel vulnerability reproduction.
Given a CVE entry with its artifacts (e.g., description, patch, PoC and references), a target kernel version, and optionally a baseline \configfile, \tool aims to (i) assess triggerability \emph{under a specified evaluation protocol} and (ii) derive a small, auditable configuration boundary.
The output is a configuration boundary $S$: a satisfiable set of \texttt{CONFIG\_*} \emph{assignments} (e.g., \texttt{y}/\texttt{m}) that remains enabled in the \emph{effective} \configfile after \makeolddef.
When a runnable reproducer (PoC) is available, building and booting the kernel with the effective \configfile and executing the reproducer should yield matched trigger evidence (e.g., a crash or a \kasan report); we then prune redundant assignments while preserving this evidence to obtain a 1-minimal triggering boundary.
\tool requires a runnable reproducer to \emph{claim} triggerability; without one, it does not claim triggerability and outputs only an effective (post-\makeolddef) configuration boundary.

\subsection{Setting and Assumptions}
Our reproduction setting targets a controlled environment built from upstream Linux kernel source code rather than executing a PoC directly on a distribution's default kernel.
The configuration context, when provided, serves as a baseline for evaluating triggerability, while the upstream-based build enables tighter control over confounding factors (e.g., heavy customization and non-upstream patches in distribution kernels).
Concretely, we generate a \configfile for the target version, compile it into a bootable kernel image (e.g., \texttt{bzImage}) via the \kbuild~\cite{kbuild} system, and boot the resulting kernel together with a root filesystem in QEMU/KVM.
We then run the PoC inside the virtualized system and use kernel logs (e.g., crash traces or \kasan reports) to determine whether matched trigger evidence is observed.
\tool aims to derive a production-tailored configuration boundary while validating it in this controlled upstream-based setting; applying the resulting boundary to heavily patched distribution kernels or hardware-specific environments may require additional adaptation that is orthogonal to the configuration-effectiveness problem studied here.

\subsection{Key Definitions}

\textbf{\kconfig, \kbuild, and Effective Configuration.}
The Linux kernel's configurable build process is governed jointly by \kconfig~\cite{kconfig} and \kbuild~\cite{kbuild}.
\kconfig specifies \texttt{CONFIG\_*} symbols and logical constraints among them, while \kbuild (realized through the Makefile hierarchy) translates whether a symbol is enabled into concrete compilation decisions, such as building code into the kernel image or as a loadable module.
Configuration constraints commonly include \texttt{depends on}, which requires condition $Y$ to hold before option $X$ can be enabled, and \texttt{select}, which forces option $Y$ to be enabled when $X$ is enabled.
We explicitly distinguish between a \emph{written configuration}, i.e., values manually set in \configfile, and an \emph{effective configuration}, i.e., the final \configfile obtained after running \makeolddef.
The \olddef step fills in default values according to \kconfig rules and may silently disable options whose prerequisites are unsatisfied; consequently, an option written as enabled does not guarantee that it remains enabled (as \texttt{y} or \texttt{m}) in the effective \configfile.
We treat an option as \emph{effective} only if (i) it retains its intended value in the effective \configfile and (ii) a subsequent build confirms that \kbuild realizes the corresponding compilation units in the expected artifact form, i.e., built into the kernel image for \texttt{=y} or produced as a \texttt{.ko} module for \texttt{=m}.

\textbf{Candidate Set and Minimal Triggering Boundary.}
We define the \emph{candidate set} $C$ as the collection of \texttt{CONFIG\_*} options inferred from vulnerability artifacts (e.g., CVE description, patch, and references) and build system files (e.g., \kconfig and \kbuild Makefiles) as potentially relevant to triggering.
The candidate set may contain redundant options, and some options may not remain enabled in the effective \configfile produced by \makeolddef.
A \emph{triggering configuration boundary} $S$ is a set of \texttt{CONFIG\_*} assignments that is satisfiable under \kconfig constraints, remains enabled in the effective \configfile after \makeolddef, and produces matched trigger evidence when the PoC is executed.
We call $S$ \emph{1-minimal} (subset-minimal) if removing any single assignment from it causes at least one of the following to fail under the same evaluation protocol: (i) the configuration becomes ineffective after \makeolddef, (ii) the kernel cannot be built/booted, or (iii) matched trigger evidence is no longer observed.
This notion of minimality is defined for an \emph{operational evaluation boundary}, where necessity is judged by buildability/bootability, effectiveness, and trigger evidence under the protocol.

\textbf{Trigger Evidence.}
After executing the PoC in our reproduction environment, we may observe concrete abnormal behaviors, such as a system crash or a \kasan report; we treat such observable abnormalities as trigger evidence.
To support stable minimization, we compare evidence using a consistent matching rule and accept an option removal only when the matched trigger evidence is still observed.
Concretely, we use a \emph{bug signature} extracted from logs (e.g., sanitizer/crash type and a normalized stack-trace pattern) to decide whether two runs correspond to the same bug.

\textbf{Evaluation Protocol.}
Under our setting, a configuration is considered reproducible if
(i) it satisfies \kconfig constraints and, after \makeolddef, the assignments in the boundary remain enabled in the effective \configfile and are realized by \kbuild in the compiled artifacts, and
(ii) building and booting the kernel with the effective \configfile and executing the PoC produces matched trigger evidence.
The concrete oracle definitions and protocol parameters (e.g., signature matching and repeated runs) are specified in Sec.~\ref{sec:oracles}, and the evaluation settings (e.g., datasets and timeouts) are specified in Sec.~\ref{sec:setup}.

\textit{Scope and Threat Model.}
\tool targets configuration prerequisites for triggering a vulnerability with a given reproducer in a controlled environment.
For downstream assessment, the inferred boundary should be interpreted as a configuration precondition, not as a standalone exploitability result. FCC targets deployments in which a production-tailored kernel exposes the relevant kernel interface to inputs beyond vendor- or administrator-controlled code, such as untrusted local users, tenant workloads, sandboxed applications, containers, guest workloads, or network/device inputs that can reach the vulnerable subsystem. If the relevant interface is reachable only by trusted vendor code, the inferred boundary should be interpreted as a configuration precondition rather than an exploitable-risk conclusion.

It does not aim to synthesize exploits, strengthen reliability beyond the stated protocol, or cover distribution-specific patch stacks and proprietary hardware dependencies.
When a vulnerability requires unavailable hardware, privileged setup, or non-upstream components, \tool may fail to validate triggerability even if the vulnerability is affected at the source level; we treat such cases as outside the scope of protocol-defined trigger validation.

\section{Method}

\subsection{Definitions and Oracles}
\label{sec:oracles}

Given a CVE with its artifacts (description, patch, external references, and an optional proof-of-concept (PoC)),
a target kernel version $v$, and an optional baseline configuration $B$ (defaulting to \texttt{defconfig}),
\tool outputs a configuration boundary.
We represent a boundary as a set of \texttt{CONFIG\_*} \emph{assignments} $S$ (e.g., forcing a symbol to \texttt{y} or \texttt{m}) applied as overrides on top of $B$.
When a runnable PoC allows establishing a reference bug signature, \tool further minimizes the boundary to a 1-minimal triggering boundary $S_{\min}$.
We write $B \oplus S$ to denote injecting $S$ into $B$ (i.e., assignments in $S$ override $B$ for the same symbol).
\textbf{Convention:} In the pseudocode, we treat an ``assignment'' as its target symbol together with an intended enabling value; adding/removing an element corresponds to enforcing/not enforcing that assignment in the boundary.

When $S_{\min}$ is defined, it satisfies the following properties under the evaluation protocol.
\textbf{(i) Effectiveness:} after injecting $S_{\min}$ into $B$ and running \makeolddef, each assignment in $S_{\min}$ remains in its intended state in the resulting effective \configfile, and the build artifacts confirm \kbuild realization of the corresponding units (built-in or module).
\textbf{(ii) Triggering:} building and booting the kernel from this effective \configfile and executing the PoC produces matched trigger evidence.
\textbf{(iii) 1-minimality (subset-minimality):} for every $s\in S_{\min}$, removing $s$ (i.e., not enforcing that assignment) causes at least one of the following to fail under the same protocol:
(a) matched trigger evidence is no longer observed,
(b) the kernel cannot be built/booted, or
(c) the boundary becomes ineffective after \makeolddef.
This minimality is operational: it is defined for a buildable, protocol-validated boundary.
Without a runnable PoC (or if no reference signature can be established), \tool does not claim triggerability and outputs only an effective (post-\makeolddef) boundary $S$.

We instantiate the above properties with two oracle functions.


\textit{Effectiveness oracle $\textsf{Effective}(B,S)$.}
We run \makeolddef on $B\oplus S$ to obtain $\textsf{Olddef}(B\oplus S)$.
The oracle returns \textsf{true} if and only if
(a) each assignment in $S$ has its expected value in the resulting effective \configfile;
and (b) the subsequent build succeeds and confirms that the retained assignments are realized by \kbuild in the artifact form implied by their effective values, i.e., built into the kernel image for \texttt{=y} or produced as loadable modules for \texttt{=m}.
Otherwise, including configuration rollback, missing expected artifacts, or build failures, it returns \textsf{false}.

\textit{Triggering oracle $\textsf{Trigger}(K,P)$.}
We execute the PoC $P$ on a booted kernel instance $K$ (built from an effective \configfile).
The oracle returns \textsf{true} only if a trigger signal is observed \emph{and} the observed evidence matches a bug signature.
Trigger signals include \texttt{panic}/\texttt{oops}, \kasan/\kmsan reports, or a crash with a consistent signature.
We extract a bug signature from kernel logs using an LLM-assisted matcher (e.g., sanitizer/crash type, top stack frames, and key subsystem/function tokens).
To reduce false positives, \tool establishes a reference signature $\Sigma_{\textit{ref}}$ from an initial triggering run; subsequent tests are considered triggering only if they match $\Sigma_{\textit{ref}}$.
If $\Sigma_{\textit{ref}}$ cannot be established (e.g., the initial run does not yield a matchable signature), \tool does not claim triggerability and falls back to outputting only an effective boundary.
Failures to boot, execute, or timeouts, as well as outputs that cannot be matched reliably, are treated as non-triggering.

To handle non-deterministic triggering, we use $\textsf{Trigger}^t$:
we repeat the execution $t$ times and deem the vulnerability triggered if at least $\tau$ runs return \textsf{true} (i.e., match $\Sigma_{\textit{ref}}$).
In this work we set $t=5$ and $\tau=1$.
Note that in this paper, ``minimal'' refers to 1-minimality (subset-minimality), not minimum-cardinality optimality.

\subsection{Overview}

\begin{figure}[t]
  \centering
  \includegraphics[width=\textwidth]{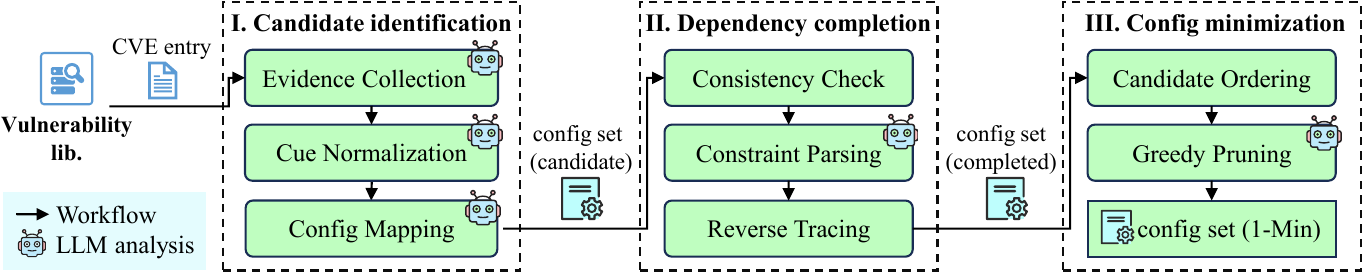}
  \caption{Overview of \tool's workflow (Stage~I: candidate identification; Stage~II: dependency completion; Stage~III: config minimization).}
  \label{overview}
\end{figure}

\tool follows the three-stage workflow in Fig.~\ref{overview} and maintains four artifacts:
vulnerability cues $L$, a candidate set $C$, an effective boundary $S$, and a dependency graph $G$.
Stage~I constructs $C$ from vulnerability artifacts and build specifications.
Stage~II completes missing prerequisites under \makeolddef feedback, yielding an effective boundary $S$ and an edge set $E$ used to build $G$.
Stage~III minimizes $S$ under the triggering oracle, guided by $G$, and outputs $S_{\min}$ when trigger validation is available.

\subsection{Candidate Identification}
Stage~I derives a candidate set $C$ related to the vulnerability's triggering conditions to constrain subsequent completion and minimization.
We start from the \nvd entry (CVE description and CWE) and incorporate an associated PoC when available.
We then parse the \nvd reference list to obtain URLs and tags (e.g., Vendor Advisory, Patch, Exploit).
For each URL, we retrieve the page and extract the main textual content by removing non-content elements (e.g., scripts/styles) and boilerplate sections (e.g., navigation, footers, tables of contents, and comments).
We retain CVE-specific information (e.g., affected components/versions, triggering conditions, and patch/mitigation details) and summarize it to control context length.
References without substantive text are excluded; if the evidence remains insufficient, we fall back to the CVE description and CWE text only.
For reproducibility, we snapshot and cache retrieved contents (raw HTML and extracted text) with access timestamps, and run cue extraction on cached snapshots.

\begin{figure}[t]
  \centering
  \includegraphics[width=\textwidth]{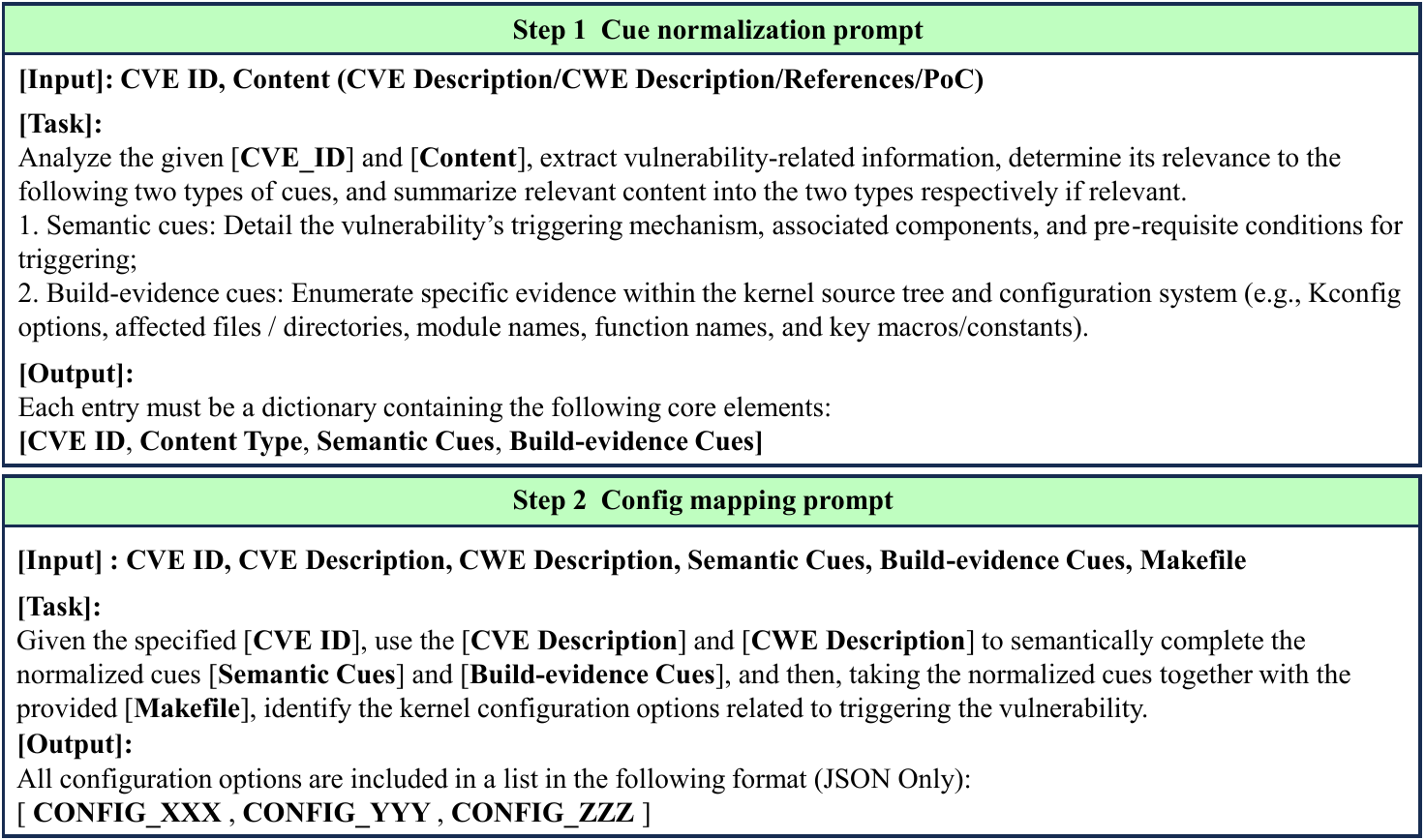}
  \caption{\tool candidate identification: a two-step prompting workflow. (1) Cue normalization extracts configuration-relevant semantic cues and build-evidence cues from heterogeneous vulnerability sources. (2) Config mapping infers kernel options by combining the normalized cues with build specifications.}
  \label{prompt}
\end{figure}

From the evidence context, we extract configuration-relevant cues and normalize them into two types:
\emph{semantic cues} describing the triggering mechanism, involved components, and preconditions; and
\emph{build-evidence cues} providing anchors such as file paths, module names, key interfaces/function names, and macros/constants.
We then construct prompts with the normalized cue set $L$ and build specifications, guiding the model to map cues to kernel configuration options and produce a candidate set $C$ (Fig.~\ref{prompt}).
We constrain outputs to a machine-parsable schema (Sec.~\ref{sec:setup}).
In Stage~I, the model is used only to propose candidates; candidates are retained or discarded by \makeolddef-based effectiveness checking and, when applicable, trigger validation.
We fix the model version and decoding parameters (e.g., temperature $=0$) and log prompts/responses for replay.

\subsection{Dependency Completion}

Stage~II addresses the ``set $\neq$ effective'' failure mode, where directly writing inferred options into \configfile can still lead to silent rollback after \makeolddef.

\textit{Consistency Check.}
\tool injects the boundary into baseline $B$ and runs \makeolddef to obtain $\textsf{Olddef}(B \oplus S)$.
If an assignment fails to reach its intended state (e.g., rolled back to \texttt{not set} or otherwise altered), it is treated as ineffective.
\tool iteratively completes prerequisites for ineffective targets until they become stably effective or a termination condition is met (empty worklist, maximum iterations, or no satisfiable branch under bounded search).

\textit{Constraint Parsing.}
For each ineffective target symbol $x$, \tool locates its definition in the \kconfig corpus and extracts constraints from:
(i) the \texttt{config} block for $x$ and its \texttt{depends on} expression,
(ii) enclosing \texttt{if}/\texttt{endif} contexts, and
(iii) conditional contexts at each \texttt{source} site along the inclusion chain that brings the definition into the build.
Backtracking stops once either
(1) all symbols referenced in the \texttt{source}-site conditions are enabled in the realized configuration, or
(2) \tool reaches the top-level \kconfig and no further enclosing \texttt{source} site exists.
To avoid redundant parsing and cyclic-inclusion overhead, \tool deduplicates visited nodes $(\textit{\kconfig file path}, \textit{source statement location})$ and merges repeated conditional constraints before completion.

\textit{Reverse Dependency Tracing.}
Many interface/aggregation symbols (e.g., \texttt{VIRTIO}) are not directly user-selectable; they are often enabled implicitly via \texttt{select} from other options.
If we only traverse forward prerequisites (\texttt{depends on}), we may miss such feasible enabling paths.
For a missing prerequisite symbol $y$ of a target $x$, \tool searches the \kconfig corpus for \texttt{select y} and collects potential selector options $\mathrm{Selectors}(y)$.
These selectors represent alternative satisfaction branches for keeping $y$ effective in the current context.
We filter branches using semantic cues and cap the retained branches to $k$ (we use $k=1$) to control branching.
Selected selector options are added to $S$, and any additional prerequisites they introduce are handled by subsequent \makeolddef-feedback iterations.

\textit{Edge Convention.}
We use a directed edge $u \rightarrow v$ to denote that keeping $u$ effective requires $v$ under the current context.
This covers both standard prerequisites induced by \texttt{depends on}/\texttt{if}/\texttt{source} and satisfaction branches induced by \texttt{select}-mediated enabling.
Algorithm~\ref{alg:completion} presents the completion procedure.
Here, $Src$ denotes the \kconfig corpus for version $v$.
\texttt{OlddefCheck} runs \makeolddef on $B\oplus S$ and returns the realized configuration and the subset of targets in $S$ that fail to keep their intended states.

\begin{algorithm}[t]
\caption{Dependency completion ($u \rightarrow v$ means keeping $u$ effective requires $v$).}
\label{alg:completion}
\small
\begin{tabularx}{\linewidth}{@{}lX@{}}
\textbf{Input:} & $C,B,L,Src$ \\
\textbf{Output:} & $S,E$ \\
\end{tabularx}
\begin{algorithmic}[1]
\State $S\leftarrow C;\;E\leftarrow\emptyset$
\State $(CfgFinal,Disabled)\leftarrow OlddefCheck(B,S)$
\State $Q\leftarrow Disabled;\;Visited\leftarrow\emptyset;\;iter\leftarrow 0$
\While{$Q\neq\emptyset$ \textbf{and} $iter<ITER\_MAX$}
 \State $iter\leftarrow iter+1$
 \State $x\leftarrow pop(Q)$; 
 \If{$x\in Visited$}
 \textbf{continue}
 \EndIf
 \State $Visited\leftarrow Visited\cup\{x\}$
 \State $Kfrag\leftarrow ExtractKconfigFragments(Src,x)$
 \State $Y_{miss}\leftarrow AnalyzeMissingDeps(CfgFinal,x,Kfrag)$
 \For{\textbf{each} $y\in Y_{miss}$}
   \State $E\leftarrow E\cup\{x\rightarrow y\}$;\; $S\leftarrow S\cup\{y\}$
   \State $Sel\leftarrow FindSelectors(Src,y)$
   \For{\textbf{each} $a\in PruneByClues(Sel,L,k)$}
     \State $E\leftarrow E\cup\{y\rightarrow a\}$;\; $S\leftarrow S\cup\{a\}$
   \EndFor
 \EndFor
 \State $(CfgFinal,DisabledNew)\leftarrow OlddefCheck(B,S)$;\; $Q\leftarrow Q\cup DisabledNew$
\EndWhile
\State \Return $(S,E)$
\end{algorithmic}
\end{algorithm}


\textit{Graph Construction.}
From the completed boundary $S$ and edge set $E$, we construct
$G=(V,A,D_{\text{in}})$ with $V=S$,
$A[u]=\{v\mid(u,v)\in E\}$, and
$D_{\text{in}}(v)=|\{u\mid(u,v)\in E\}|$.
Thus, $D_{\text{in}}(v)$ is the number of current assignments that
depend on $v$; isolated assignments remain in $V$ with dependent
count zero.

\subsection{Config Minimization}

\begin{figure}[t]
  \centering
  \includegraphics[width=\textwidth]{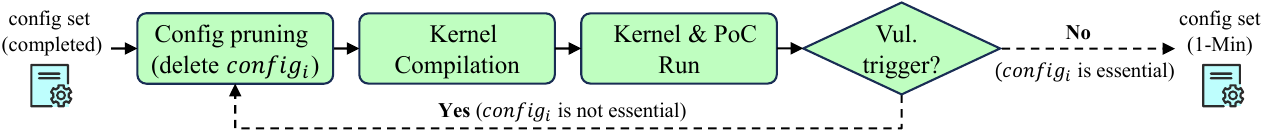}
  \caption{Runtime-feedback-driven configuration minimization workflow.}
  \label{fig:minimization}
\end{figure}

Stage~III removes redundant assignments while preserving matched trigger evidence.
We use a runtime feedback loop as shown in Fig.~\ref{fig:minimization}: for each candidate assignment $c$, we tentatively remove it, re-run \makeolddef, rebuild and boot the kernel, and execute the PoC, judging triggerability using $\textsf{Trigger}^t$ (Sec.~\ref{sec:oracles}).
If build/boot fails, or the run no longer matches the reference signature, $c$ is treated as non-removable.
We prioritize candidates with $D_{\text{in}}(c)=0$ (no dependents) to reduce cascading deactivation; each candidate is attempted at most once.
If $c$ is necessary, we freeze its prerequisite closure to preserve satisfiability and operational 1-minimality.

Algorithm~\ref{alg:min} instantiates this loop using $G=(V,A,D_{\text{in}})$.
It initializes a worklist $Q$ with candidates that have no dependents, and maintains:
\emph{Tried}, to attempt each candidate at most once, and
\emph{Frozen}, to prevent removing assignments in the prerequisite closure of a necessary one.
$\textsf{IsRedundant}(c,S,P)$ implements the try-remove test described above.
If $c$ is removed, we decrement the dependent counts of its prerequisites $v\in A[c]$ and push newly-unblocked prerequisites with $D_{\text{in}}(v)=0$ into $Q$.
If $c$ is necessary, \textsf{PropagateNecessity} freezes $c$'s prerequisite closure along outgoing edges.

\begin{algorithm}[!ht]
\caption{Config Minimization (1-minimal)}
\label{alg:min}
\small
\begin{tabularx}{\linewidth}{@{}lX@{}}
\textbf{Input:} & $G=(V,A,D_{\text{in}}),S,P$ \\
\textbf{Output:} & $S_{\min}$ \\
\end{tabularx}
\begin{algorithmic}[1]
\State $Q\leftarrow\{v\mid D_{\text{in}}(v)=0\}$;\; $Tried\leftarrow\emptyset$;\; $Frozen\leftarrow\emptyset$
\While{$Q\neq\emptyset$}
 \State $c\leftarrow pop(Q)$
 \State \textbf{if } \(c\in Tried \lor c\in Frozen\)
        \textbf{ then continue}
 \State $Tried\leftarrow Tried\cup\{c\}$
 \If{$IsRedundant(c,S,P)$}
   \State $S\leftarrow S\setminus\{c\}$
   \For{\textbf{each} $v\in A[c]$}
     \State $D_{\text{in}}(v)\leftarrow D_{\text{in}}(v)-1$
     \If{$D_{\text{in}}(v)=0$}\State $Q\leftarrow Q\cup\{v\}$\EndIf
   \EndFor
 \Else
   \State $PropagateNecessity(c,Frozen,G)$
 \EndIf
\EndWhile
\State \Return $S_{\min}\leftarrow S$
\end{algorithmic}
\end{algorithm}

\section{Experimental Setup}
\label{sec:setup}
We conduct experiments to evaluate \tool and answer the following research questions (RQs).

\begin{itemize}

  \item \textbf{RQ1: Configuration identification and redundancy.}
  Can \tool identify a vulnerability-triggering candidate option set while introducing as few irrelevant options as possible?
  We compare \tool with \kernjc~\cite{Ruan2024RAID} in terms of candidate-set redundancy and coverage of the available reference option sets.
  
  \item \textbf{RQ2: Correctness of dependency completion.}
  When candidate options become ineffective due to unmet dependencies, can \tool recursively infer and add the required prerequisites so that critical options remain enabled in the final \configfile?
  We evaluate \tool's ability to complete dependencies for critical options on two datasets~\cite{KernelCTF2026,Ruan2024RAID}.

  \item \textbf{RQ3: Effectiveness of runtime-validated minimization.}
  While preserving PoC triggerability, can \tool remove redundant configuration options through runtime validation?
  We compare configuration size before and after minimization, record the number of pruning attempts, and verify that the minimized configuration can still trigger the vulnerability.

  \item \textbf{RQ4: Efficiency of FCC stage-wise execution.}
  What is the resource cost of FCC across its three stages, and which stage dominates the end-to-end execution cost? We measure token consumption and execution time for candidate identification, dependency completion, and runtime-validated minimization. We further evaluate a CVE-focused evidence-selection variant of Stage I to quantify the reducible overhead in the dominant stage.

\end{itemize}

\subsection{Evaluation Datasets}
We evaluate \tool on the \kernjc dataset~\cite{kernjcdb} and the KernelCTF dataset~\cite{KernelCTF2026}, both containing Linux kernel CVE samples.
The \kernjc dataset includes proof-of-concept (PoC) programs for 66 CVEs, together with the configuration option sets identified by \kernjc.
The KernelCTF dataset contains 57 CVEs, and each CVE's description document specifies the configuration options that must be additionally enabled beyond the default configuration.
We refer to such options required on top of the default configuration as \emph{critical configuration options}.

Since \tool is designed to assess and synthesize vulnerability-triggering configurations under production-tailored settings, our evaluation focuses on vulnerabilities that require enabling additional options beyond the default configuration.
To this end, we manually reproduced all 66 CVEs in the \kernjc dataset and found that 34 can be successfully triggered under the default configuration generated by \makedef without enabling any extra options.
These CVEs do not exercise \tool's ability to identify and complete critical configuration options and are therefore excluded from our evaluation set.
As a result, we retain the remaining 32 CVEs from the \kernjc dataset that require additional configuration beyond \texttt{defconfig}.

\kernjc and KernelCTF overlap on one CVE.
After removing this overlap from KernelCTF, our final evaluation dataset consists of 88 CVEs (32 from \kernjc and 56 from KernelCTF).


We use this 88-CVE dataset to evaluate \tool's capability in configuration identification and dependency completion (RQ1 and RQ2).  
For RQ1, KernelCTF provides document-level required options and is used as the external reference for coverage evaluation. 
For \kernjc, which does not provide document-level ground truth for critical configuration options, we evaluate whether the initial candidate sets contain the runtime-validated trigger boundary and how much redundancy they introduce. 
We further perform a \kernjc-start minimization control to check whether minimizing the \kernjc-provided sets leads to the same trigger-preserving boundary as \tool. RQ3 is conducted on the 32 \kernjc CVEs with runnable PoCs.

\subsection{Baseline}
\textbf{Baseline for RQ1.}
\kernjc~\cite{Ruan2024RAID} is a representative approach for Linux kernel CVE configuration analysis.
It infers vulnerability-related options primarily from \kconfig dependency relations and provides identified configuration sets for the CVEs in its dataset, making it suitable for comparison.

\textbf{Baseline for RQ2.}
For the \emph{configuration injection} step, we use an \emph{olddef-only} baseline strategy:
after injecting candidate options into \configfile, we run \makeolddef once to normalize the configuration under \kconfig constraints and obtain the final effective configuration used for building.
This baseline captures the standard kernel build behavior where options with unmet dependencies may be silently disabled during configuration normalization.

\subsection{Evaluation Metrics}
We report evaluation metrics aligned with the four research questions.

\textbf{Configuration Identification (RQ1).}
We evaluate identification quality from two perspectives: coverage of critical configuration options and redundancy of the inferred candidate set.
Let $C$ denote the candidate set inferred by a method (\tool or \kernjc), and let $K$ denote the critical option set when document-level critical-option labels are available.
We define a CVE as \emph{covered} if all critical options are included in the candidate set (i.e., $K \subseteq C$).
Accordingly, we report \emph{CVE Coverage (\%)} as the percentage of CVEs in the dataset that are covered.
To measure redundancy, we report the \emph{Average Candidate Set Size} (Avg.\ $|C|$).
Since the goal is to encompass $K$ with as few irrelevant options as possible, a smaller $|C|$ (under high coverage) indicates lower redundancy.
For KernelCTF CVEs, $K$ is taken from the accompanying documents (i.e., options that must be enabled beyond \texttt{defconfig}).
For the \kernjc dataset, the provided configuration sets contain redundant options and do not include document-level ground-truth critical configuration options. 
We therefore use the runtime-validated trigger boundary as the reference set for RQ1. 
This boundary is first obtained by \tool's minimization and then cross-checked by a \kernjc-start minimization control: starting from each \kernjc-provided set, removing the options outside this boundary still preserves effectiveness and the same trigger signature. 
Thus, on \kernjc, CVE coverage measures whether a candidate set contains the experimentally confirmed trigger boundary, while Avg.\ $|C|$ measures the redundancy introduced to contain that boundary.

\textbf{Dependency Completion (RQ2).}
Dependency completion is evaluated based on two dimensions: effectiveness after configuration normalization and efficiency of the inference process.
For each CVE, we inject the completed configuration into \configfile, run \makeolddef, and verify whether \emph{all} critical options in $K$ remain enabled in the final configuration.
We report:
(i) the \emph{configuration success rate}, defined as the percentage of test cases (CVEs) for which all critical options take effect without rollback; and
(ii) the \emph{convergence efficiency}, measured by the number of inference iterations required for \tool to reach a stable configuration state.

\textbf{Runtime-Validated Minimization (RQ3).}
We evaluate minimization efficiency and convergence using structural and procedural statistics of the minimization process.
Specifically, we report
(i) the configuration reduction rate, measured by the number of options before and after minimization ($|C|$ vs.\ $|S_{\min}|$);
(ii) the size of the dependency graph used in minimization, including the number of nodes $|V|$ and edges $|A|$; and
(iii) the total number of pruning rounds until convergence.
We additionally verify that $S_{\min}$ preserves PoC triggerability under the trigger oracle.

 \textbf{Runtime and Token Overhead (RQ4).} We measure per-CVE wall-clock time and LLM token use for each FCC stage. Stage I covers evidence collection, cue normalization, and candidate identification; Stage II covers dependency completion and repeated make olddefconfig checks; Stage III covers runtime-validated minimization, including kernel compilation, VM booting, and PoC execution. Because Stage III requires runnable PoCs, we report it on the KernJC subset; for KernelCTF, we report Stage~I and Stage~II overhead only.

\subsection{Implementation Details} \label{sec:impl}
We report the implementation and environment details required for exact replay, including the toolchain/kernel build settings, oracle parameters (timeouts and iteration limits), and the fixed LLM version/decoding configuration used in Stage~I.
The FCC prototype is implemented in approximately 3,200 lines of Python code, leveraging the LangChain framework for orchestration.

FCC fetches CVE pages and external references, extracts plain text, and caches snapshots for replay. We use OpenAI GPT-5.1 (snapshot: gpt-5.1-2025-11-13) with temperature 0. The model is invoked for Stage~I candidate identification, Stage~II constraint parsing, and Stage~III trigger-evidence matching. We enforce JSON outputs with schema/regex checks and retry invalid responses, flagging persistent failures for manual inspection.

For consistency and fairness in comparison with \kernjc, we strictly adopt the same target kernel versions as used in \kernjc for experiments on the \kernjc dataset.
For the KernelCTF dataset, we additionally apply \kernjc's version identification procedure to determine the corresponding target kernel version for each CVE.
This ensures that \tool and \kernjc perform configuration inference, dependency completion, and trigger validation under the same (or equivalent) version context, avoiding biases introduced by kernel-version discrepancies.

Stage~III runs on a 144-core Intel Xeon Platinum 8352V server with \SI{500}{\giga\byte} RAM. We compile the Stage~II completed configuration, boot the kernel with a BusyBox rootfs~\cite{Busybox} under QEMU/KVM~\cite{Bellard05USENIX}, and run the PoC inside the VM. Experiments use Ubuntu~20.04.6 LTS; kernels~\(\leq\)~4.x are built in an Ubuntu~16.04.7 Docker container to avoid toolchain incompatibilities.

FCC validates \kbuild realization by mapping each retained option to its \kbuild Makefile entry and checking the expected output. For example, \texttt{CONFIG\_NF\_TABLES} maps to \texttt{obj-\$(CONFIG\_NF\_}\allowbreak\texttt{TABLES)} \texttt{+=} \texttt{nf\_tables.o}; \texttt{=y} requires the object in \texttt{built-in.a}/\texttt{built-in.o}, and \texttt{=m} requires a \texttt{.ko} listed in \texttt{modules.order}. FCC records the matched entry and observed output as audit evidence.

To ensure deterministic evaluation and prevent resource exhaustion, we enforced strict iteration limits and timeouts throughout the pipeline.
Specifically, during the dependency completion process in Stage~II, we set the maximum number of iterations to 10 to effectively preclude infinite loops during dependency inference.
For the configuration minimization phase in Stage~III, we imposed explicit timeout thresholds for critical operations: the maximum time allowed for kernel compilation is set to \SI{300}{\second}, while both virtual machine booting and PoC execution are strictly limited to \SI{60}{\second}.
Any operation exceeding these time windows is deemed a failure for the current iteration.

\section{Experimental Results}
\label{sec:results}

\subsection{Configuration Identification and Redundancy (RQ1)}
Table~\ref{tab:rq1-coverage-size} presents the performance of \tool and \kernjc in configuration identification on two evaluation datasets (\kernjc and KernelCTF).
On the \kernjc dataset (32 CVEs), both \tool and \kernjc successfully covered all 32 CVEs under the runtime-validated trigger-boundary reference.
\tool achieves full coverage with substantially fewer candidate options: 471 candidates in total (Avg.\ 14.72 per CVE), compared to 2208 candidates for \kernjc (Avg.\ 69.00 per CVE).
To check whether this reference boundary is specific to \tool's starting set, we perform a \kernjc-start minimization control. 
For each \kernjc configuration set, we remove all options outside \tool's validated \(S_{\min}\), rerun \texttt{make olddefconfig}, rebuild and boot the target kernel, and re-execute the PoC. 
The resulting configurations still pass \texttt{make olddefconfig}, build and boot successfully, and reproduce the same trigger signature for 32 CVEs. 
This confirms that \kernjc and \tool reach the same trigger-preserving boundary on this subset, while \tool starts from a far smaller set.

\begin{table}[t]
\caption{Configuration identification performance of \tool and \kernjc on two evaluation datasets. ``Covered CVEs'' counts the number of CVEs whose critical option set $K$ is fully included in the candidate set $C$. For KernelCTF, the reference set is taken from documented required options; for \kernjc, it is the runtime-validated trigger boundary cross-checked by the \kernjc-start minimization control.}
\label{tab:rq1-coverage-size}
\centering
\small
\setlength{\tabcolsep}{6pt}
\begin{tabular}{lccccc}
\toprule
\textbf{Dataset} & \textbf{\#CVEs }& \textbf{Method} & \textbf{Covered CVEs} & \textbf{Avg.\ $|C|$ (Total)} & \textbf{Coverage (\%)} \\
\midrule
\multirow{2}{*}{\kernjc}
& \multirow{2}{*}{32}
& \tool   & 32 & 14.72 (471)    & 100.0 \\
&        & \kernjc & 32 & 69.00 (2208)   & 100.0 \\
\midrule
\multirow{2}{*}{KernelCTF}
& \multirow{2}{*}{56}
& \tool   & 53 & 27.51 (1458)   & 94.6 \\
&        & \kernjc & 50 & 201.76 (10088) & 89.3 \\
\bottomrule
\end{tabular}
\end{table}

On the KernelCTF dataset (56 CVEs), \tool covers 53 CVEs (94.6\%), while \kernjc covers 50 CVEs (89.3\%).
\tool produces a much smaller candidate set (1458 in total; Avg.\ 27.51 per CVE) than \kernjc (10088 in total; Avg.\ 201.76 per CVE), indicating lower redundancy while maintaining high coverage.

\begin{figure}[t]
  \centering
  \includegraphics[width=\textwidth]{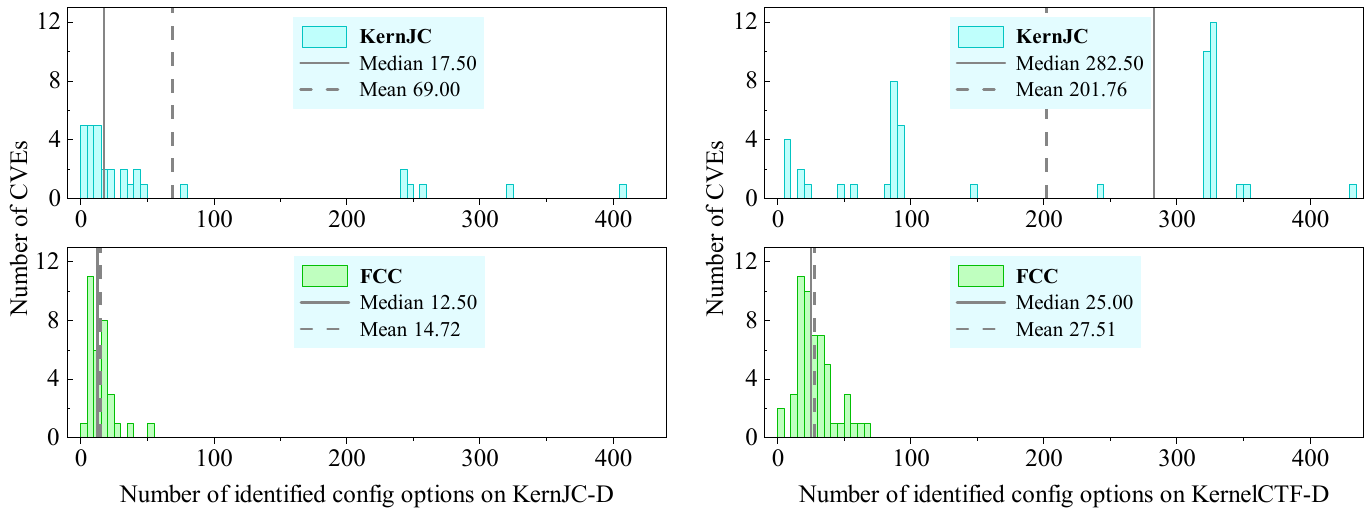}
  \caption{Comparison of FCC and KernJC in the number of identified configuration options on the KernelCTF and KernJC datasets.}
  \label{rq1_fig}
\end{figure}

Figure~\ref{rq1_fig} shows the frequency distribution of the numbers of identified candidate options, highlighting structural differences between the two methods. FCC exhibits a highly concentrated, left-skewed distribution on both datasets, with most CVEs falling in the low-count range, indicating consistently smaller candidate sets. In contrast, KernJC shows a much broader dispersion with a pronounced long tail, especially on the KernelCTF dataset, where candidate counts vary substantially and extend to high-redundancy regimes. Overall, the figure complements Table~\ref{tab:rq1-coverage-size} by showing that FCC not only reduces redundancy on average, but also does so consistently across CVEs while maintaining high coverage.

\begin{tcolorbox}
\textbf{Answer to RQ1.}
\tool achieves higher coverage than \kernjc on the KernelCTF dataset while producing substantially smaller candidate sets, reducing redundancy for subsequent dependency completion and minimization. On the \kernjc dataset, the \kernjc-start control reaches the same runtime-confirmed boundary, showing that \tool's smaller start set (14.72 vs. 69.00 options/CVE) removes redundancy without losing post-\texttt{make olddefconfig}, build/boot success, or trigger evidence.
\end{tcolorbox}

\subsection{Correctness of Dependency Completion (RQ2)}
To quantify the effectiveness of \tool in resolving configuration dependency issues, we compare \tool against the olddef-only baseline strategy (running \makeolddef after injection without additional inference) across all 88 CVEs.

\begin{figure}[t]
  \centering
  \includegraphics[width=\linewidth]{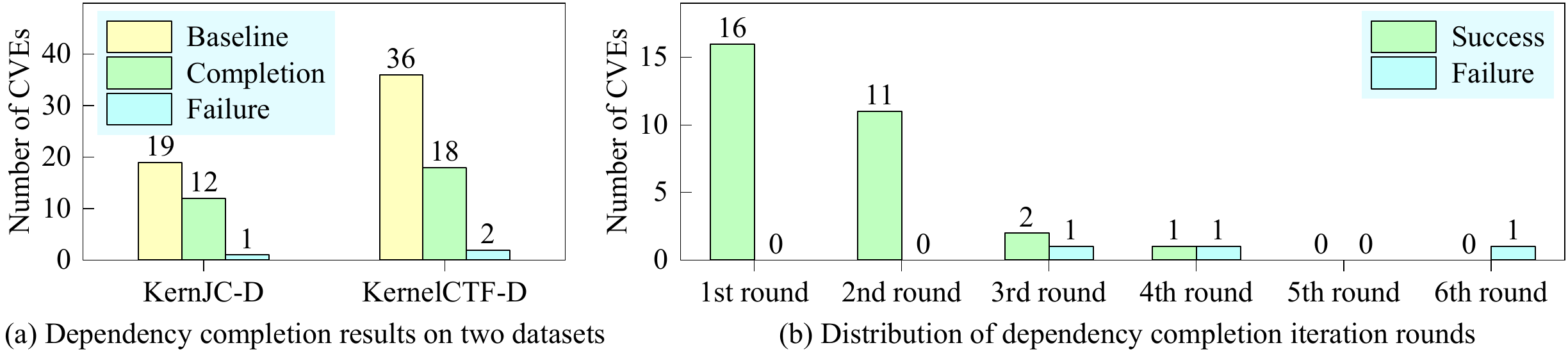}
  \caption{Effectiveness of dependency completion.}
  \label{rq2_fig1}
\end{figure}

The baseline shows clear limitations in handling implicit constraints:
out of 88 test cases, 33 CVEs (37.5\%) encounter configuration rollback, where critical options are disabled due to unsatisfied dependencies.
Figure~\ref{rq2_fig1} illustrates the breakdown of configuration outcomes and the improvement achieved by \tool.
Among these 33 baseline-failed cases, \tool resolves 30 (90.9\%) through recursive dependency inference, ensuring that all critical options remain effective in the final build.
Overall, the configuration success rate improves from 62.5\% (55/88) to 96.6\% (85/88).

We investigate the remaining 3 cases that do not fully take effect.
These failures stem from hard environmental constraints rather than defects in the dependency inference algorithm.
For instance, attempting to enable architecture-specific options (e.g., \texttt{CONFIG\_CPU\_32v7}, which is ARM-only) in an x86 reproduction environment is rejected by \kconfig.
After excluding such hard cross-architecture or hardware constraints, \tool demonstrates high efficacy in dependency repair within the reachable software configuration space.

Beyond the final success rate, we analyze the number of inference iterations required for \tool to reach a stable configuration.
Across the 33 baseline-failed cases, 48.5\% (16/33) are resolved within a single inference iteration round, and 33.3\% (11/33) within two iteration rounds, indicating that most missing dependencies are shallow.
For deeper dependency chains (e.g., CVE-2018-6555, which involves four layers of dependencies), \tool achieves stepwise completion through multiple iteration rounds.
Notably, the unresolved cases typically undergo more exploratory iterations (4.3 rounds on average, up to 6), reflecting that \tool traverses candidate dependency branches until reaching unsatisfiable constraints or exhausting the bounded search space.

\begin{tcolorbox}
\textbf{Answer to RQ2.}
\tool improves the configuration success rate from 62.5\% to 96.6\% compared to the olddef-only baseline, while converging efficiently in practice on our dataset.
\end{tcolorbox}

\begin{table}[!htbp]
  \centering
  \caption{Detailed statistics of runtime-validated configuration minimization. \textbf{PR}: Number of pruning rounds. $|S_{\min}|$ denotes the size of the 1-minimal configuration boundary.}
  \label{tab:rq3}
  \scriptsize
  \setlength{\tabcolsep}{7pt} 
  \renewcommand{\arraystretch}{1.00} %
  
  \begin{tabular}{lcccc|lcccc}
    \toprule
    \textbf{CVE ID} & \textbf{\#Nodes} & \textbf{\#Edges} & \textbf{\#PR} & \textbf{$|S_{\min}|$}  & 
    \textbf{CVE ID} & \textbf{\#Nodes} & \textbf{\#Edges} & \textbf{\#PR} & \textbf{$|S_{\min}|$}  \\
    \midrule
    
    CVE-2016-10150 & 21 & 0  & 21 & 2  & CVE-2020-28941 & 5  & 0  & 5  & 2  \\
    CVE-2016-4557  & 7  & 0  & 7  & 1  & CVE-2020-8835  & 6  & 0  & 6  & 1  \\
    CVE-2016-6187  & 10 & 0  & 10 & 1  & CVE-2021-22555 & 15 & 6  & 14 & 2  \\
    CVE-2017-16995 & 13 & 0  & 13 & 1  & CVE-2021-26708 & 14 & 15 & 12 & 4  \\
    CVE-2017-18344 & 14 & 3  & 13 & 2  & CVE-2021-27365 & 13 & 11 & 13 & 2  \\
    CVE-2017-2636  & 7  & 0  & 7  & 1  & CVE-2021-34866 & 10 & 0  & 10 & 1  \\
    CVE-2017-6074  & 13 & 0  & 13 & 1  & CVE-2021-3490  & 20 & 5  & 20 & 1  \\
    CVE-2017-8824  & 7  & 0  & 7  & 1  & CVE-2021-3573  & 5  & 0  & 5  & 3  \\
    CVE-2018-12233 & 4  & 0  & 4  & 1  & CVE-2021-42008 & 12 & 0  & 12 & 3  \\
    CVE-2018-5333  & 11 & 7  & 10 & 1  & CVE-2021-43267 & 10 & 0  & 10 & 1  \\
    CVE-2018-6555  & 33 & 38 & 32 & 2  & CVE-2022-0995  & 8  & 0  & 8  & 1  \\
    CVE-2019-6974  & 20 & 19 & 19 & 2  & CVE-2022-1015  & 12 & 0  & 12 & 2  \\
    CVE-2020-16119 & 5  & 0  & 5  & 1  & CVE-2022-25636 & 8  & 0  & 8  & 3  \\
    CVE-2020-25669 & 6  & 0  & 6  & 1  & CVE-2022-32250 & 21 & 6  & 20 & 2  \\
    CVE-2020-27194 & 7  & 0  & 7  & 1  & CVE-2022-34918 & 42 & 0  & 42 & 1  \\
    CVE-2020-27830 & 6  & 4  & 5  & 2  & CVE-2023-32233 & 12 & 0  & 12 & 2  \\
    \bottomrule
  \end{tabular}
\end{table}

\subsection{Effectiveness of Runtime-Validated Minimization (RQ3)}

Prior to configuration minimization, we assessed the reliability of the trigger oracle described in Section~\ref{sec:oracles}. For each CVE, we launched the test environment five times and, in each launch, recorded the number of PoC executions required to trigger the vulnerability; we report the maximum count observed across the five launches. The results indicate high reproducibility: 31 out of 32 CVEs are triggered by the first PoC execution in all five launches. Accordingly, to reduce false negatives due to runtime flakiness, we set the oracle parameters to $t=5$ and $\tau=1$ (as defined in Section~\ref{sec:oracles}).

Table~\ref{tab:rq3} presents detailed statistics of the minimization process for each CVE.
Here, \textbf{Nodes} denotes the number of configuration options in the effective set (after dependency completion) that must be explicitly enabled beyond the default configuration, while \textbf{Edges} represents the necessary dependency relations maintained to ensure these options remain effective.

On this dataset, the dependency topology constructed by \tool is structurally sparse.
Correspondingly, the number of pruning rounds is typically close to the number of Nodes, suggesting near-linear behavior in practice under \tool's dependency-guided pruning order.

We also quantify the size reduction before and after minimization.
The candidate configuration sets generated by \tool in the initial stage contain 14.72 options on average.
Following pruning via the runtime feedback loop, the minimized configuration sets contain 1.63 options on average, corresponding to a configuration reduction rate of 88.93\%.
Compared to \kernjc's average of 69.00 options on the same CVEs, \tool yields a much more compact configuration boundary, with an average size equal to approximately 2.36\% of the baseline size.

This reduction also contextualizes FCC against fuzzing-oriented configurations. Syzkaller's Linux-kernel configuration guide recommends feature-rich options for coverage, diagnostics, and bug detection~\cite{syzkaller_config}, and links to a syzbot reference configuration~\cite{syzbot_refconfig}. We use that linked reference configuration (\texttt{upstream-apparmor-kasan.config}) only as a scale reference: it contains 4,206 enabled \texttt{CONFIG\_*} entries set to \texttt{y} or \texttt{m}. Such configurations are appropriate for bug discovery but too broad for identifying the options required for a known CVE to remain triggerable. FCC instead prunes the CVE-derived effective set to a 1-minimal boundary, reaching 1.63 trigger-preserving options per CVE on average.

Finally, in all 32 test cases, the minimized configuration sets pass validation under the trigger Oracle:
the kernel built from the minimized configuration can still execute the PoC and reproduce the expected crash or error report under the same protocol.
This indicates that \tool's ``try-remove-validate'' loop converges to an \emph{oracle-level} 1-minimal set without breaking triggerability on our dataset.
We discuss threats to validity and generalization considerations in Sec.~\ref{sec:threats}. 

\begin{tcolorbox}
\textbf{Answer to RQ3.}
\tool's topology-guided runtime minimization removes redundant configuration options while preserving vulnerability triggerability under the evaluation protocol.
On the \kernjc CVEs, \tool achieves an 88.93\% reduction on average (14.72 to 1.63 options), yielding a compact 1-minimal evaluation boundary.
\end{tcolorbox}

\subsection{Efficiency of FCC Stage-Wise Execution (RQ4)}

\begin{figure}[H]
  \centering
  \includegraphics[width=\linewidth]{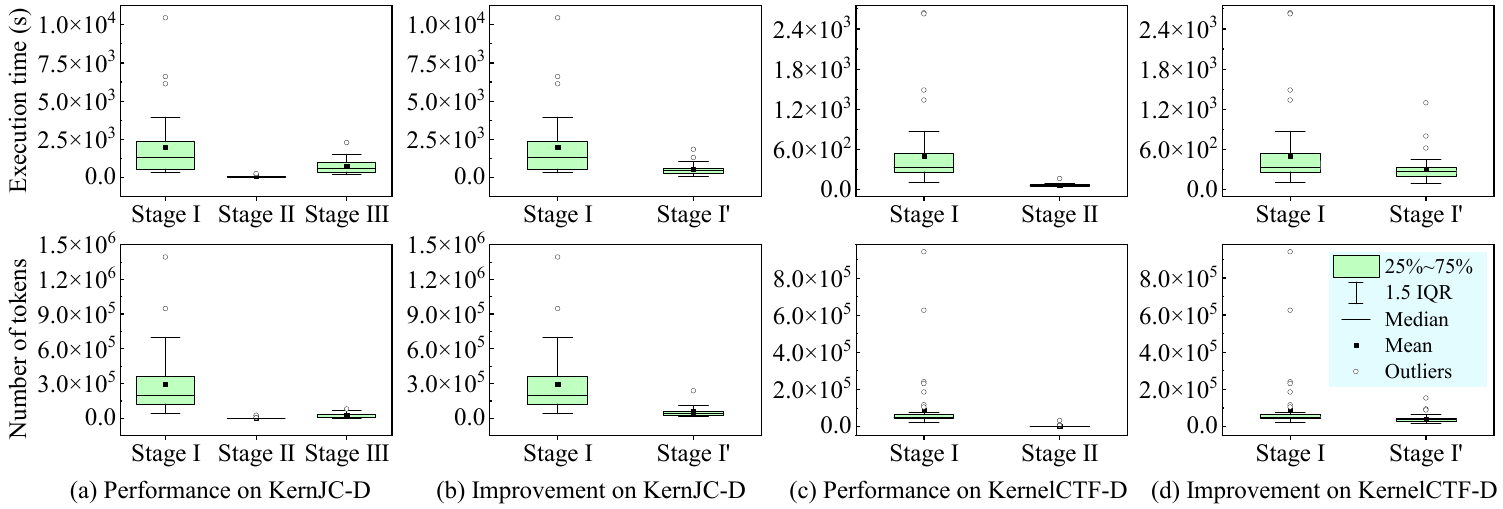}
  \caption{Runtime and token overhead of FCC. Stage I$'$ denotes the CVE-focused evidence-selection variant of Stage I.}
  \label{rq4}
\end{figure}

Figure~\ref{rq4} reports per-CVE time and token costs. Stage~I dominates FCC's overhead: \SI{1977}{\second} / 291.8k tokens on KernJC-D and \SI{502}{\second} / 86.6k tokens on KernelCTF-D, because it processes noisy CVE evidence before candidate mapping. Stage~I$'$ pre-filters retrieved references before prompting, without using validation outcomes: it keeps pages whose extracted text contains target-CVE evidence and excludes generic or non-substantive pages that inflate context length. Candidate mapping and all deterministic validations remain unchanged, so Stage~I$'$ measures avoidable evidence-processing overhead rather than a different inference/validation pipeline. Stage~I$'$ reduces costs to \SI{532}{\second} / 56.2k tokens on KernJC-D and \SI{304}{\second} / 40.4k tokens on KernelCTF-D, i.e., 26.9\% / 19.3\% and 60.5\% / 46.6\% of the original time/token costs, respectively.

Stage~II remains lightweight, taking \SI{68}{\second} / 2.1k tokens per CVE on KernJC-D and \SI{72}{\second} / 1.4k tokens on KernelCTF-D. 
Stage~III is reported on KernJC-D because it requires runnable PoCs; its per-CVE cost of \SI{717}{\second} mainly reflects repeated build/boot/PoC runs, while its LLM token use of 21.6k tokens is attributable to trigger-evidence matching. Across Stage~III validation runs, rebuilds and reboots take \SI{91}{\second} and \SI{5}{\second} on average, respectively.

\begin{tcolorbox}
\textbf{Answer to RQ4.}
Stage I is the main overhead source and can be substantially reduced by CVE-focused evidence selection. Stage II remains lightweight, while Stage III's cost reflects the repeated build/boot/PoC validation needed to preserve runtime triggerability.
\end{tcolorbox}

\section{Discussion}
\subsection{Using the 1-Minimal Boundary for Production Assessment}

FCC’s 1-minimal boundary should be used as a configuration-level trigger predicate, not as an exact symbol checklist or a standalone exploitability claim. A production assessment should normalize the vendor configuration on the corresponding kernel tree (e.g., via make olddefconfig or a vendor-equivalent process) and compare against the effective configuration. The comparison is role-aware: core options guarding the vulnerable subsystem, interface, driver, or compilation unit must remain effective and Kbuild-realized, whereas dependency-enabling options are witnesses and may be replaced by equivalent selector paths. For example, a kernel that realizes \texttt{CONFIG\_VIRTIO} through \texttt{CONFIG\_VIRTIO\_MMIO} can satisfy the same dependency even if FCC retained \texttt{CONFIG\_VIRTIO\_PCI}.

This conclusion remains configuration-level. Vendor patches, architecture/hardware constraints, and deployment reachability may still prevent runtime triggering; with a runnable PoC, triggerability should be confirmed by matched trigger evidence on the production-effective configuration. Otherwise, FCC reports configuration-level reachability only.

\subsection{Error Analysis of Incomplete Cases}

Table~\ref{tab:error_analysis} reports symbol-level omissions or rejected branches, not CVE-level trigger failures. The three Stage~I omissions are recoverable because dependency completion can still reach the missed symbols from broader Stage~I anchors, such as CIFS, UNIX, or VSOCKETS; if such anchors are also missed, FCC may not recover the prerequisite. In contrast, the three rejected Stage~II branches are guarded by architecture- or hardware-specific constraints, such as ARM-only or PMU/KVM-dependent paths, and cannot remain effective in our x86 setting. FCC therefore rejects them after make olddefconfig and Kbuild validation rather than treating them as ordinary missing dependencies.

\begin{table}[H]
\centering
\caption{Symbol-level error analysis of incomplete or rejected branches.}
\label{tab:error_analysis}
\small
\setlength{\tabcolsep}{3pt}
\renewcommand{\arraystretch}{1.05}
\resizebox{\linewidth}{!}{%
\begin{tabular}{l|l|l|l|l}
\hline
\textbf{Type} & \textbf{CVE ID} & \textbf{Symbol} & \textbf{Error source} & \textbf{FCC outcome} \\ \hline
Stage~I omission & CVE-2023-5345 & CONFIG\_SMBFS & SMB3 evidence mapped to CONFIG\_CIFS & Recovered via CONFIG\_CIFS \\ \hline
Stage~I omission & CVE-2024-36972 & CONFIG\_AF\_UNIX\_OOB & OOB logic hidden behind AF\_UNIX & Recovered via CONFIG\_UNIX \\ \hline
Stage~I omission & CVE-2025-21756 & CONFIG\_VSOCKETS\_LOOPBACK & Loopback transport not explicit in PoC & Recovered via CONFIG\_VSOCKETS \\ \hline
Stage~II constraint & CVE-2019-6974 & CONFIG\_KVM\_ARM\_HOST & KVM/VFIO/IOMMU cues & Rejected as ARM-only under x86 \\ \hline
Stage~II constraint & CVE-2024-26808 & CONFIG\_VDSO & vDSO artifacts in PoC & Rejected due to ARMv7 dependency \\ \hline
Stage~II constraint & CVE-2023-6931 & CONFIG\_HW\_PERF\_EVENTS & perf\_event\_open() usage & Rejected as ARCH/PMU-specific path \\ \hline
\end{tabular}%
}
\end{table}

\section{Threats to Validity}\label{sec:threats}

\textit{Internal Validity.}
\textbf{Stability of LLM-based inference.} Stage~I uses an LLM to parse unstructured descriptions; output variability may change the initial candidate set $C$ (RQ1). Mitigation: we fix the LLM version and use temperature 0, and treat LLM outputs only as candidates---all options are deterministically validated by \kconfig constraints and \makeolddef, which filters hallucinations before they enter the final buildable configuration. 

\textbf{Scope of Kconfig value types.} FCC focuses on boolean/tristate reachability, consistent with Kconfig’s select semantics. Non-boolean \texttt{int}, \texttt{hex}, and \texttt{string} symbols are inherited from the baseline and make olddefconfig unless vulnerability evidence explicitly constrains their values; in our 88-CVE benchmark, no minimized boundary required such a constraint.
\textbf{Reliability of the trigger oracle.} Kernel reproducers can be flaky (especially races), so oracle errors may either over-prune necessary options or retain redundant ones (RQ3). Mitigation: we use a repeated-run oracle $\textsf{Trigger}^t$ ($t=5$, $\tau=1$) with signature matching; however, this does not guarantee \emph{oracle-level} 1-minimality under extremely low-probability triggers.

\textit{External Validity.}
\textbf{Dataset representativeness.} We evaluate on 88 CVEs from \kernjc and KernelCTF, which may not cover all subsystems or proprietary drivers. We exclude 34 \texttt{defconfig}-trivial cases and release the CVE list and filters to make the scope explicit; conclusions mainly apply to upstream vulnerabilities with public reproduction artifacts and are limited when only text descriptions are available.

\textbf{Generalizability to downstream kernels.} FCC is validated on upstream kernels. Applying an inferred boundary to distribution or vendor-patched trees still requires vendor-side adaptation: normalizing the production configuration on the downstream tree, mapping vendor-only options or patches to the inferred boundary, and confirming interface reachability. Triggerability also remains conditional on the required architecture, hardware, and virtualization environment.
\textbf{Hardware and environmental constraints.} The remaining 3 RQ2 failures are environment-limited (e.g., arch-specific options unavailable on our infrastructure). \tool does not emulate missing peripherals or cross-architecture setups; therefore, triggerability is conditional on the required virtualization/hardware environment.

\section{Related Work}
\label{sec:related}

Vulnerability reproduction often fails due to missing or mismatched context~\cite{Mu2018USENIX}.
Configuration prerequisites are often under-specified and hard to trace from artifacts~\cite{Chen2021Trustcom}; for kernels, even automated environment generation remains configuration-sensitive~\cite{Ruan2024RAID}.
We position \tool on \kconfig/\kbuild-aware reasoning, outputting an \emph{effective} (post-\makeolddef) boundary and, with a PoC, an \emph{operational 1-minimal} triggering boundary.

\textit{Vulnerability reproducibility and missing reproduction context.}
Reports frequently omit reproduction-critical details and can be internally inconsistent; vulnerability databases may also contain noisy affected-version information~\cite{Mu2018USENIX,DingIST2025,Nguyen16ESE}.
Prior work either recovers coarse configuration-related evidence from artifacts~\cite{Chen2021Trustcom} or refines affected-version ranges using commit/code evidence~\cite{Bao22ICSE}.
In contrast, \tool synthesizes prerequisites that survive \makeolddef and supports protocol-guided minimization when trigger evidence exists.

\textit{Automation of reproduction and cross-environment adaptation.}
PoC/exploit adaptation and analysis pipelines mainly target portability across kernels/versions or ecosystems~\cite{Zou2024NDSS,Jiang2023SP,Bao2017SP,Pitigalaarachchi23CSS,Lukas24USENIX}.
\tool instead keeps the reproducer fixed and infers configuration prerequisites that may be missing or become ineffective after \kconfig/\kbuild constraint resolution.

\textit{Kernel configuration inference and \kconfig/\kbuild-aware analysis.}
KernJC expands static \kconfig dependencies to identify vulnerability-related options~\cite{Ruan2024RAID}.
Related work studies traceability/documentation~\cite{Ahmmed24ASE}, scalable sampling~\cite{Fernandez-Amoros23ASE}, specification defects~\cite{oh2021FSE}, conflict resolution~\cite{Franz21ICSE}, and variability bugs~\cite{Abal17TOSEM}.
\tool adds post-\makeolddef effectiveness/\kbuild realization checks and, with a PoC, derives a runtime-validated 1-minimal boundary under a fixed protocol.

\textit{LLMs for reproduction and debugging assistance.}
LLM-based systems generate tests/steps and use feedback for reproduction (e.g., LIBRO~\cite{kang23ICSE}, AdbGPT~\cite{feng2024ICSE}, CrashTranslator~\cite{Huang24ICSE}, ReBL~\cite{Wang24ISSTA}) and can be improved via self-debugging training (e.g., LeDex~\cite{Jiang24NURIPS}).
In contrast, \tool confines LLMs to candidate identification, constraint parsing, and evidence matching, then gates them by \makeolddef-based effectiveness checks and protocol-defined trigger validation.

\section{Conclusion}
\label{sec:conclusion}

This paper targets trigger-configuration inference for Linux kernel CVEs under production-tailored settings, where upstream reproducers often omit configuration context and static dependency expansion is prone to rollback under \makeolddef and over-approximation. We presented FCC, which mines configuration candidates from vulnerability evidence, completes \kconfig prerequisites under \makeolddef feedback, and prunes options using runtime trigger evidence. FCC outputs a configuration boundary that is effective after \makeolddef and 1-minimal under an oracle-defined protocol validated in our upstream QEMU/KVM setting. On 88 historical CVEs, FCC improves over \kernjc (RQ1) and an olddef-only injection baseline (RQ2): the post-\makeolddef effectiveness rate increases from 62.5\% to 96.6\%, and the average candidate set size on the KernJC dataset drops from 69.00 to 14.72 while retaining coverage. On the 32 KernJC CVEs with runnable PoCs, FCC further reduces the option set from 14.72 to 1.63 on average without changing the matched trigger signature under the same protocol (RQ3). 
Stage-wise analysis identifies Stage~I as the main overhead source and CVE-focused evidence selection as an effective way to reduce this cost (RQ4).
FCC claims triggerability only when a runnable PoC is available, and applying the inferred boundaries to heavily patched downstream kernels or hardware-specific environments may require additional adaptation.

\section{Data Availability}
All experimental data and source code used in this paper are available at \url{https://github.com/Cyber-wei10/FCC}; an accompanying artifact is archived on Zenodo~\cite{fcc-artifact}.

\begin{acks}
We would like to thank the anonymous reviewers for their insightful and detailed feedback. This work is funded by Basic Research Program of Jiangsu (BK20252040, BK20251414).
\end{acks}

\bibliographystyle{ACM-Reference-Format}
\bibliography{ref}


\end{document}